\documentclass{PoS}
\newcommand{\tn}[1]{\textnormal{#1}}
\newcommand{\eq}[2]{\begin{equation}\label{#1} #2 \end{equation}}

\newcommand{\TeV}[0]{\tn{ TeV}}

\usepackage{subfigure}

\title{Charged Jets in Minimum Bias p-Pb Collisions at $\sqrt{s_\mathrm{NN}} = 5.02$ TeV with ALICE}

\ShortTitle{Charged Jets in p-Pb}

\author{\speaker{R\"udiger Haake} for the ALICE collaboration\\
        Westf\"alische Wilhelms-Universit\"at M\"unster, Germany\\
        E-mail: \email{ruediger.haake@uni-muenster.de}}

\abstract{
Highly energetic jets are sensitive probes for the kinematics and the topology of nuclear collisions. Jets are produced in an early stage of the collision from hard-scattered partons, which fragment into a spray of charged and neutral particles. The measurement of jet spectra in p-Pb collisions provides an important way to quantify the effects of cold nuclear matter on jet production, fragmentation and hadronization. This is possible because the hot, dense medium produced in Pb-Pb collisions is not expected to form. Proton-Lead collisions also provide an important constraint for the nuclear parton density functions. The exact evaluation of the background from the underlying event is an important ingredient to correct the measured jet spectra. The system size in p-Pb collisions is much smaller than in Pb-Pb so that the methods for background estimation need to be refined. The analysis reported here is performed on p-Pb data taken at $\sqrt{s_\mathrm{NN}} = 5.02$ TeV by the ALICE detector at the LHC in the beginning of 2013. The focus of our analysis lies on the minimum bias charged jet spectra and their comparison to the spectra from pp collisions. For this analysis various estimates for the background and its fluctuations have been tested in p-Pb and PYTHIA simulations.}

\FullConference{The European Physical Society Conference on High Energy Physics -EPS-HEP2013\\
		18-24 July 2013\\
		Stockholm, Sweden}

\begin{document}

\section{Introduction}
Jets \cite{JetRecoInALICE} can conceptually be described as the final state produced by parton fragmentation after a hard scattering. Therefore, jets are an excellent tool to access the early stage of the collision. Jets are also a good tool for studies on initial state nuclear effects in p-Pb collisions since the early collision stage is dominated by initial state effects. Besides studying these nuclear effects, analyses of p-Pb collisions can also help with the interpretation of results obtained from Pb-Pb collisions at the same energy by serving as a baseline containing only cold nuclear matter effects. In contrast to Pb-Pb collisions, the formation of a quark-gluon plasma (QGP) is not expected in p-Pb collisions and the results from these collisions can thus be used to disentangle the cold from the hot nuclear matter effects.\\

The reconstructed jet observable is defined by the jet finding algorithm used to clusterize the particles into jets. For the presented analyses, the FastJet \cite{FastJet} package and the anti-$k_\mathrm{T}$ algorithm \cite{AntiKT} was used as the signal jet reconstruction algorithm. In addition to the jet finding algorithm the correction techniques that are applied to the measured spectra are also part of the definition of jets. To understand the effect of the background and its fluctuations, detector effects and the unfolding algorithm, several parameters and methods were chosen and evaluated to estimate the systematic uncertainty on the jet spectra.

A description of these correction techniques together with the fully corrected jet spectra and the study of their nuclear modification will be presented in this paper.\\

\section{Data analysis}
The data used for this analysis was taken during the p-Pb run at $\sqrt{s_\mathrm{NN}} = 5.02 \TeV$ in the beginning of 2013 with the ALICE experiment. The \textit{minimum bias} events used are selected by demanding at least one hit in both of the scintillator trigger detectors (V0A and V0C). In total 100M minimum bias events have been used for this analysis. The clusterized charged particles are measured as tracks by the Inner Tracking System (ITS) and the Time Projection Chamber (TPC). Tracks with $p_\mathrm{T} > 0.150 \mathrm{~GeV}/c$ and within a pseudorapidity interval $|\eta|<0.9$ were used.\\
The following corrections were applied to the raw charged jet spectra:
\begin{itemize}
  \item Background: Subtract underlying event density to correct for tracks not coming from the hard collision.
  \item Background fluctuations: Account for within-event fluctuations of the background density.
  \item Detector effects: Account for reconstruction efficiencies and momentum resolution.
\end{itemize}

While the background subtraction is a large correction to the jet momentum in Pb-Pb collisions (the density is $130 \mathrm{~GeV}/c$ per unit area in central collisions), the background density in p-Pb collisions is about a factor of 100 smaller. Nevertheless, the $p_\mathrm{T}$ of tracks produced in soft processes is not negligible and several estimates for the background have been evaluated.

\begin{figure}[ht]
\subfigure[Uncorrected background density distribution.\newline The mean value is roughly 1 GeV/\textit{c}.]{%
  \includegraphics[width=0.515\linewidth]{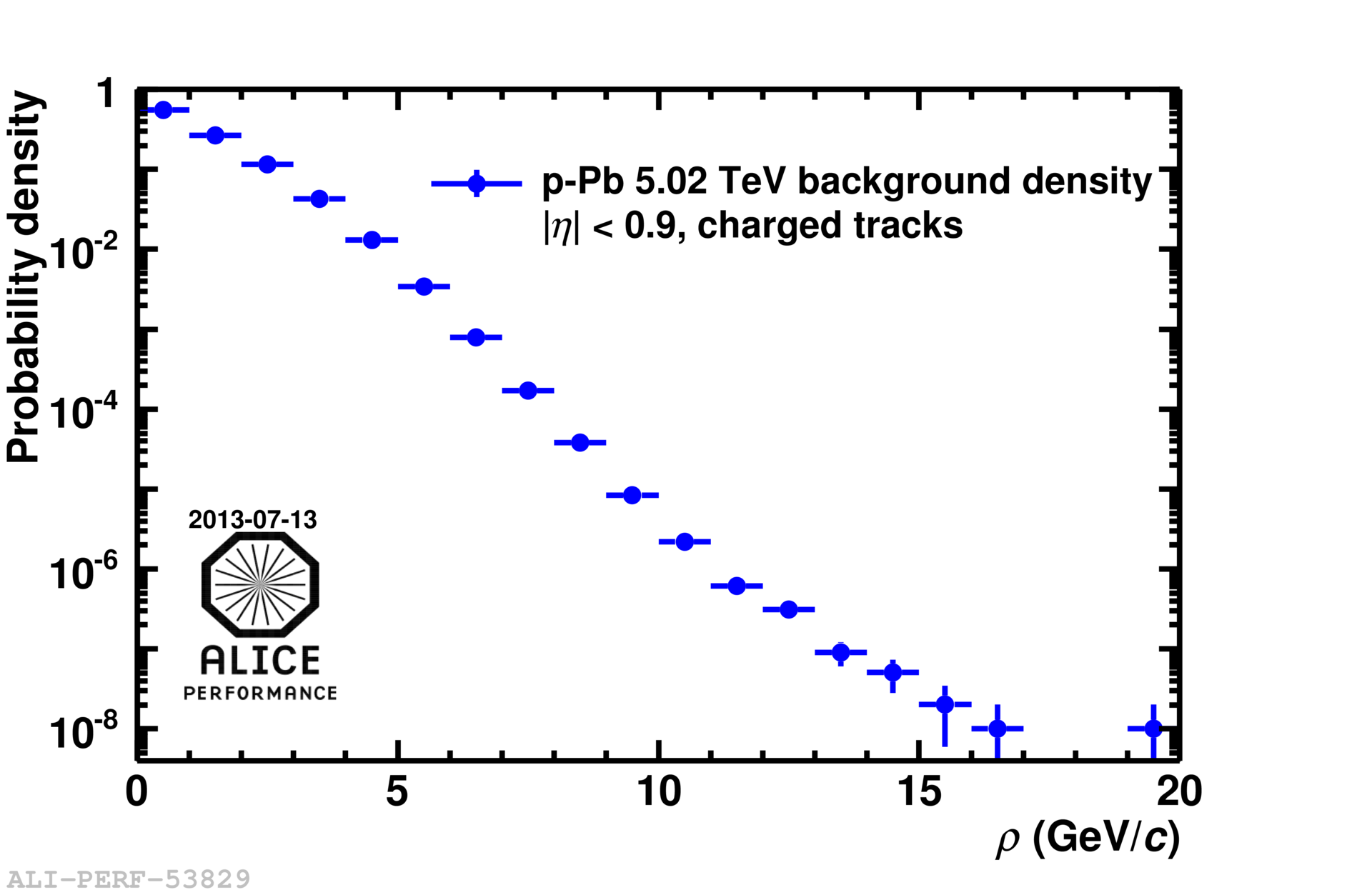} }\hfill
\quad
\subfigure[Uncorrected $\delta p_\mathrm{T}$ distribution. The variance sigma of\newline the distribution is roughly 1 GeV/\textit{c}.]{%
  \includegraphics[width=0.515\linewidth]{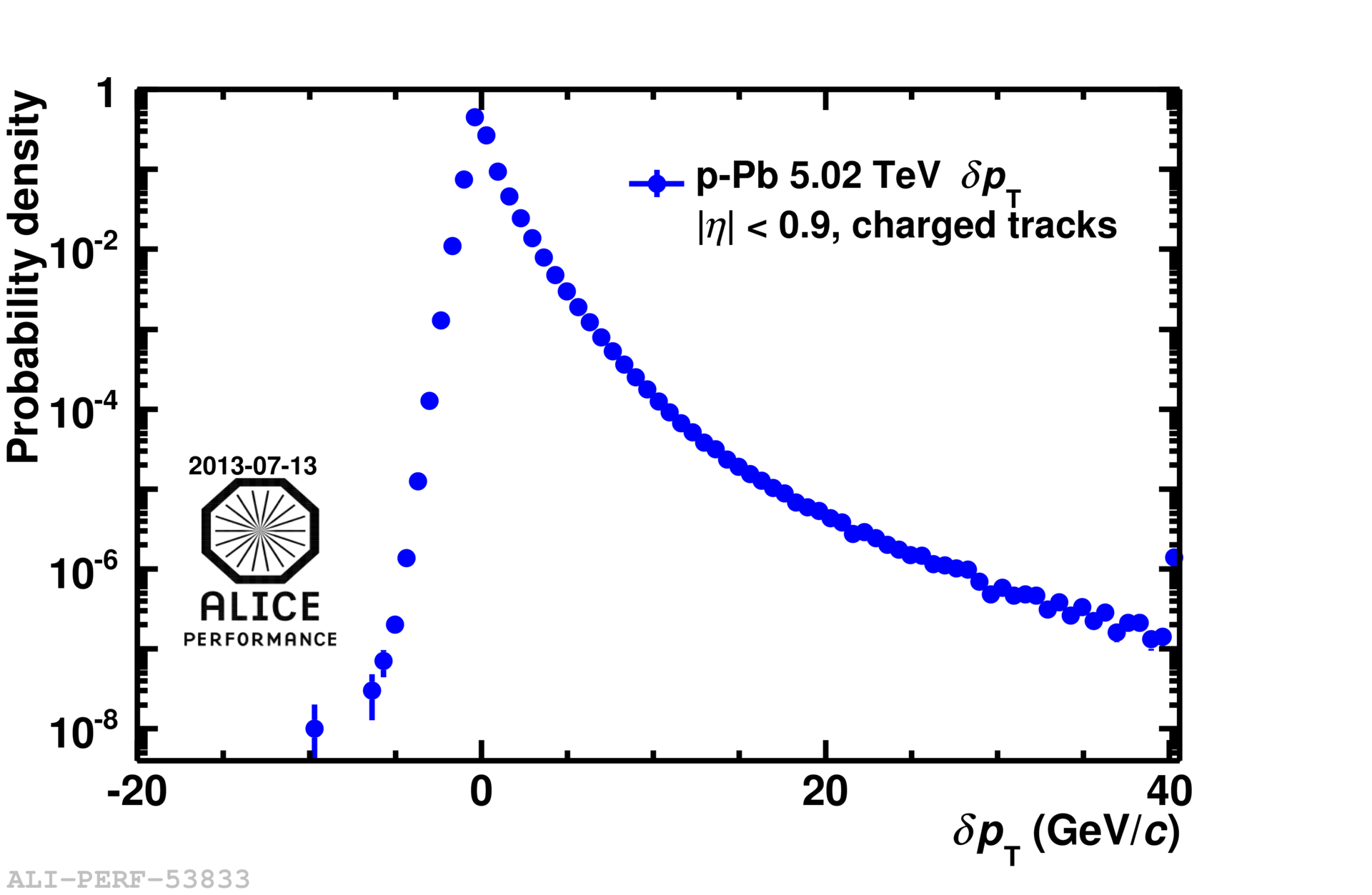} }

\caption{Probability distributions for the event-by-event background and its fluctuations calculated with the random cone approach (resolution parameter $R=0.4$).}

\label{BackgroundPlots}
\end{figure}

The correction (see figure \ref{BackgroundPlots}a) is done by calculating the background event-by-event and subtracting it from every jet. As the default approach for the background calculation we use a similar ansatz as in reference \cite{KTBackgroundCMS}. First, the $k_\mathrm{T}$ algorithm is used to clusterize the tracks of the event. This algorithm is more susceptible to contributions from soft collisions and thus to the background than the anti-$k_\mathrm{T}$ algorithm. Second, the median of the $p_\mathrm{T}$ densities of all accepted $k_\mathrm{T}$ jets is calculated. The median is used because of its stability against fluctuations and outliers. The last step is the multiplication with a correction factor $C$ that accounts for the event occupancy, which is important in sparse p-Pb collisions. Finally, the background density is given by the following equation:
\eq{def:CMSBackground}{\rho = \mathrm{median} \left\{ \frac{p_{\mathrm{T},i}}{A_i} \right\}  \cdot C,}
where $A_i$ represents the jet area and $C$ is given by
\eq{def:BackgroundOccupancy}{C = \frac{\tn{Area of all } k_\mathrm{T} \tn{ jets containing tracks}}{\tn{Acceptance}}.}
The denominator in the last equation is the detector acceptance that was clustered by the jet finding algorithm.
In contrast to the original approach in reference \cite{KTBackgroundCMS}, we only accept those $k_\mathrm{T}$ jets that do not have an overlap with the signal jets.\\

This and the other methods studied for the background calculation provide an event-by-event value for the background density. However, the real background is not equally distributed over the $\eta-\phi$ plane. The default approach is to consider these fluctuations in a statistical way by calculating the fluctuations for the full event sample and unfolding the measured distributions.\\
The distribution of the regional fluctuations around the mean background density value (see figure \ref{BackgroundPlots}b) is evaluated by the \textit{Random Cone} (RC) approach where the $\delta p_\mathrm{T}$ distribution \cite{JetBackgroundFluctuations} is calculated event-by-event by randomly throwing a cone with the same radius as used for the jet finding algorithm into the acceptance and evaluating the energy density of the tracks in this cone. The values of the $\delta p_\mathrm{T}$-distribution are therefore given by the following equation:
\eq{def:DeltaPt}{\delta p_\mathrm{T} = \sum_\mathrm{RC}{p_\mathrm{T}-\rho A}, ~~~ A = \pi R^2.}
The last correction takes the detector efficiency and resolution into account. For example, the limited efficiency causes less particles to be clusterized into a jet. To estimate this effect, a full detector simulation with PYTHIA jets and Geant3 particle transport is done. A jet response matrix is then created by geometrically matching two jet collections -- one on particle the other on detector level.\\
The latter two correction steps, correcting background fluctuations and detector effects, are applied in the unfolding procedure. For the shown results, the \textit{Singular Value Decomposition} \cite{SVDAlgorithm} method was used. Bayesian and $\chi^2$ unfolding were used for systematic comparisons and checks.\\

\begin{figure}[!htp]
\centering
\includegraphics[width=0.55\textwidth]{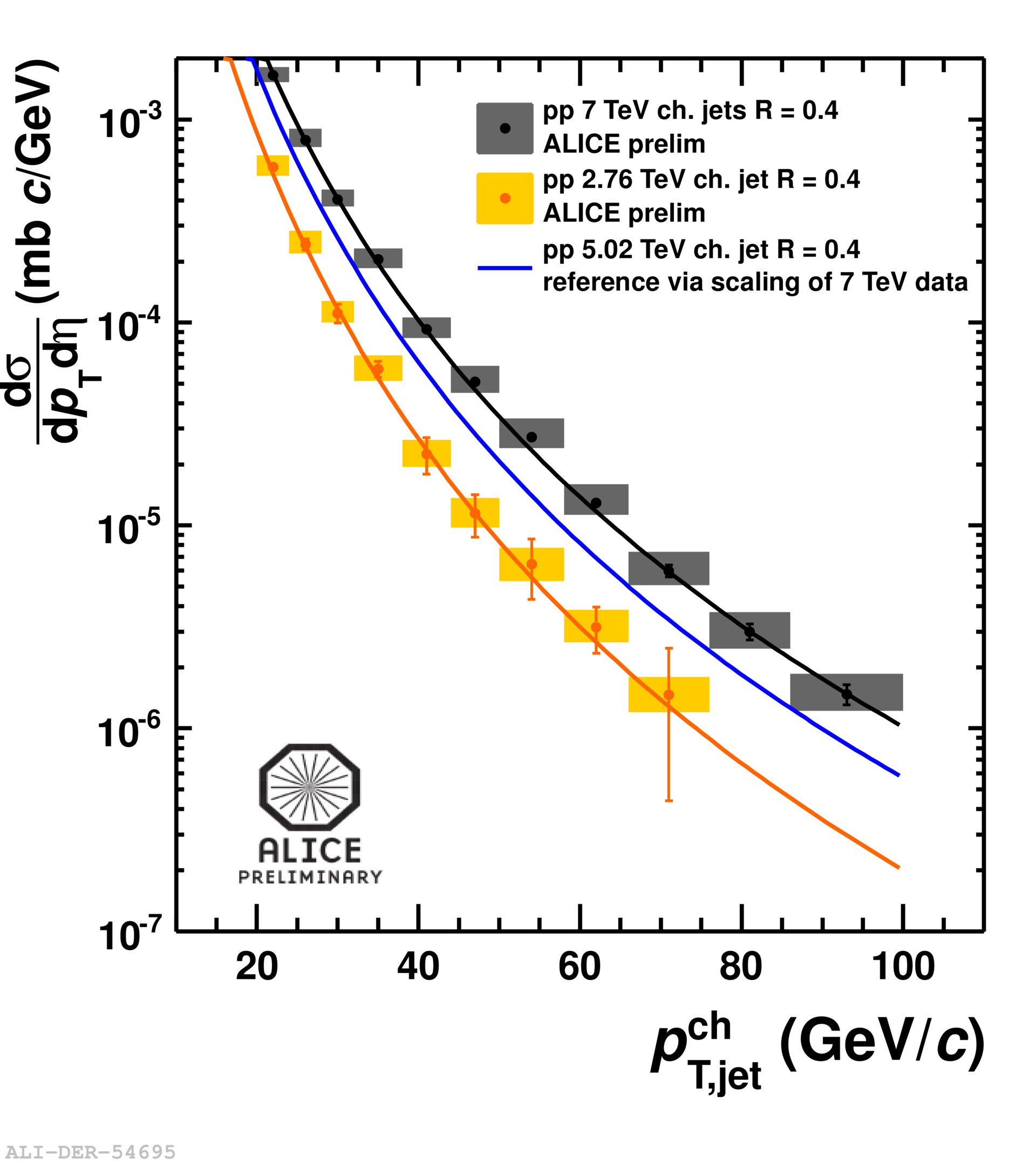}
\caption{Illustration of the scaling of the charged pp jet reference for $R_\mathrm{pPb}$. The lines show a power-law fit through the data points and the blue line is the scaled reference.}
\label{ReferencePlot}
\end{figure}

In order to quantify cold nuclear matter effects on jet production, a pp reference at the same energy as for the p-Pb spectra is needed. The pp reference was created by scaling the measured spectrum from 7 TeV using a bin-by-bin scaling factor calculated with PYTHIA. This is necessary since there is no pp data available at this energy. The scaling factor is given by the following equation: 
\eq{def:ppRefScaling}{p = \frac{\mathrm{yield}(5.02 \TeV)}{\mathrm{yield}(7 \TeV)}.}
The systematic uncertainty of the reference is $15\%$, which includes the uncertainty from the scaling method and the uncertainty on the 7 TeV spectrum itself. The scaled reference spectrum is shown together with the measured spectra at 2.76 and 7 TeV in figure \ref{ReferencePlot}.

\section{Results}
\begin{figure}[!htp]
\centering
\subfigure{\includegraphics[width=0.94\textwidth]{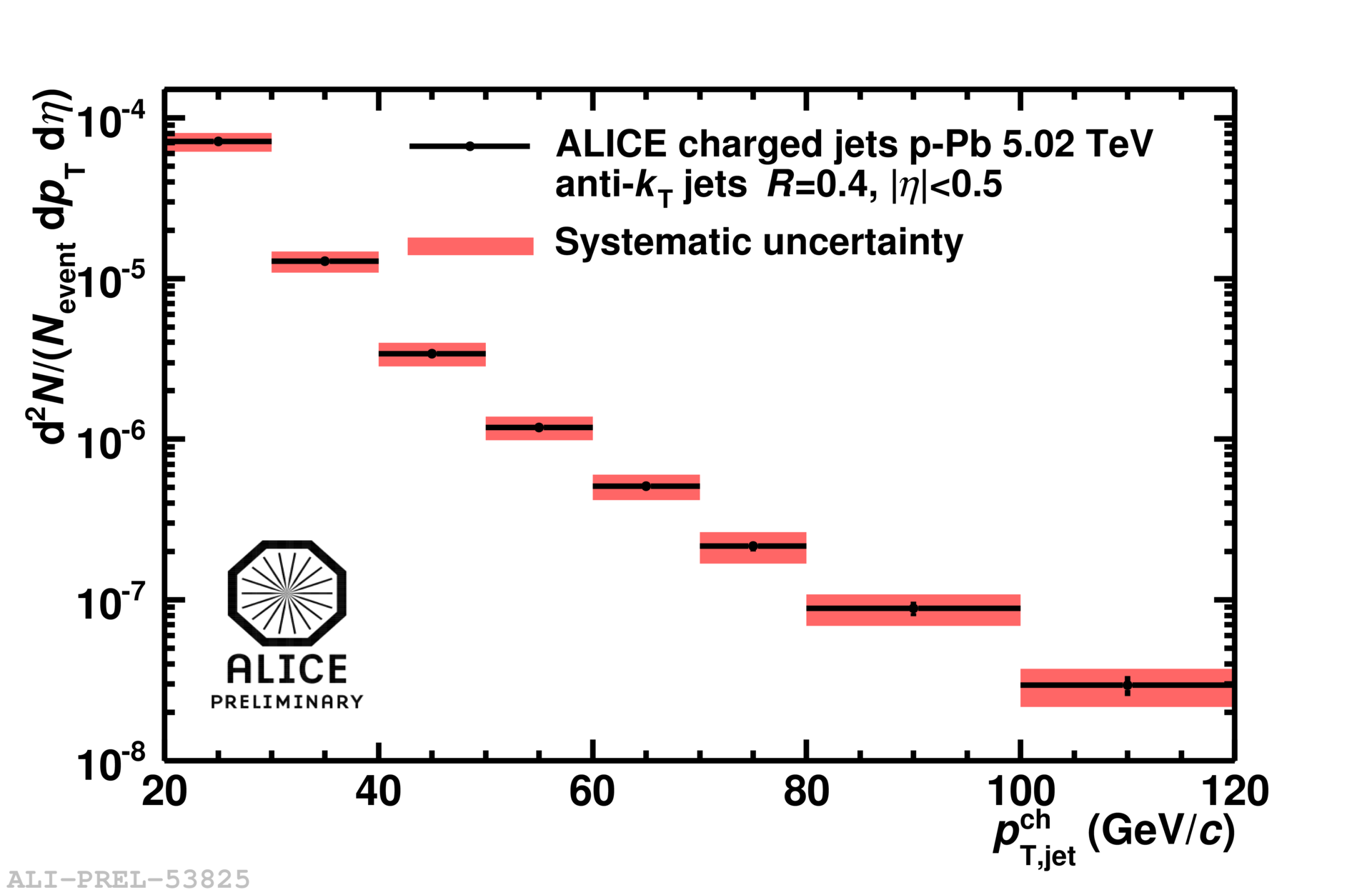}}\hfill
\caption{Charged jet spectrum at $5.02$ TeV in p-Pb collisions.}
\label{FinalYield}
\end{figure}

\begin{figure}[!htp]
\centering
\subfigure{\includegraphics[width=0.96\textwidth]{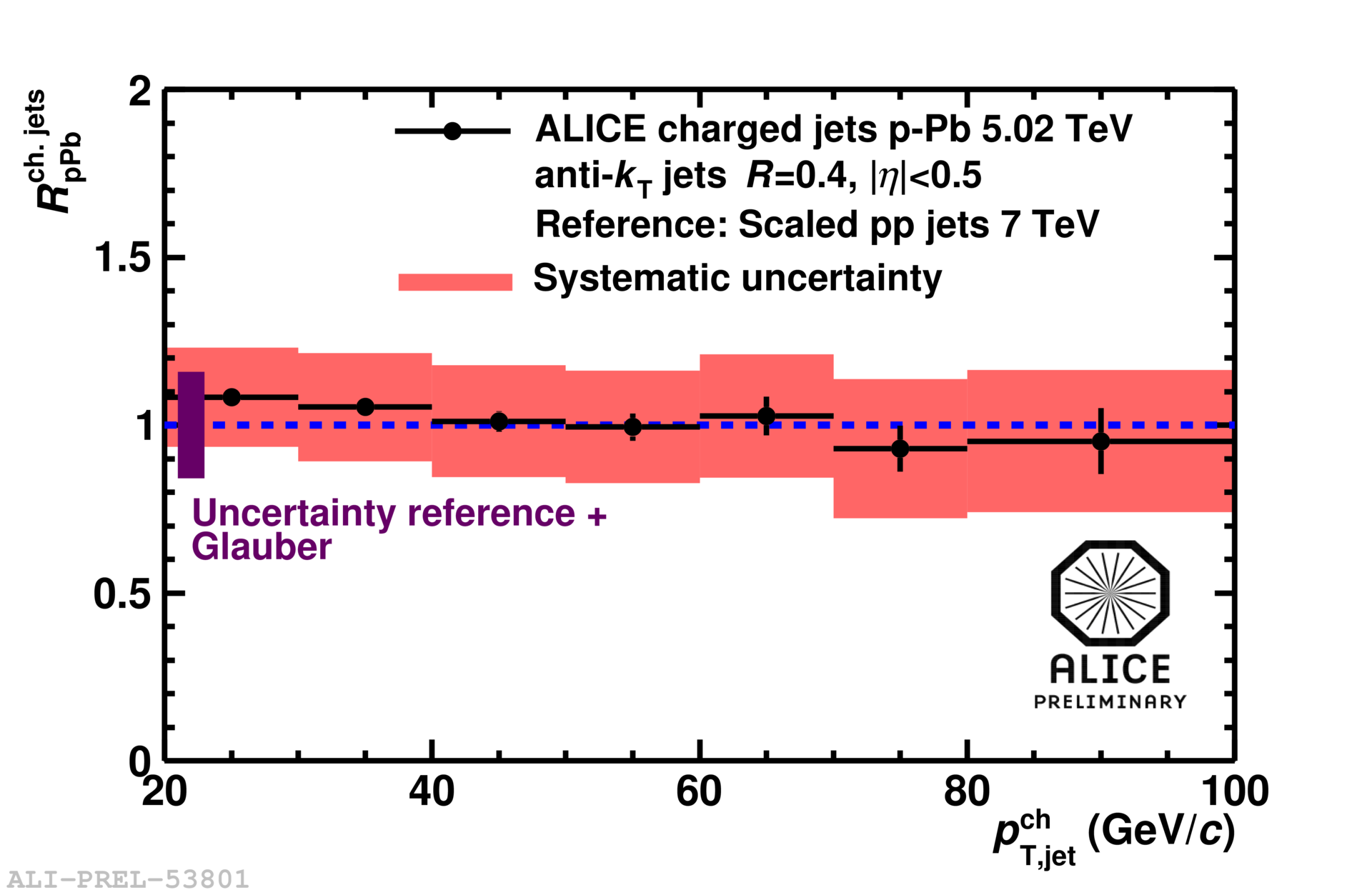}}\hfill
\caption{Nuclear modification factor of charged jets $R_\mathrm{pPb}$ with uncertainties. The systematic uncertainties are highly correlated.}
\label{FinalRpPb}

\subfigure{\includegraphics[width=0.96\textwidth]{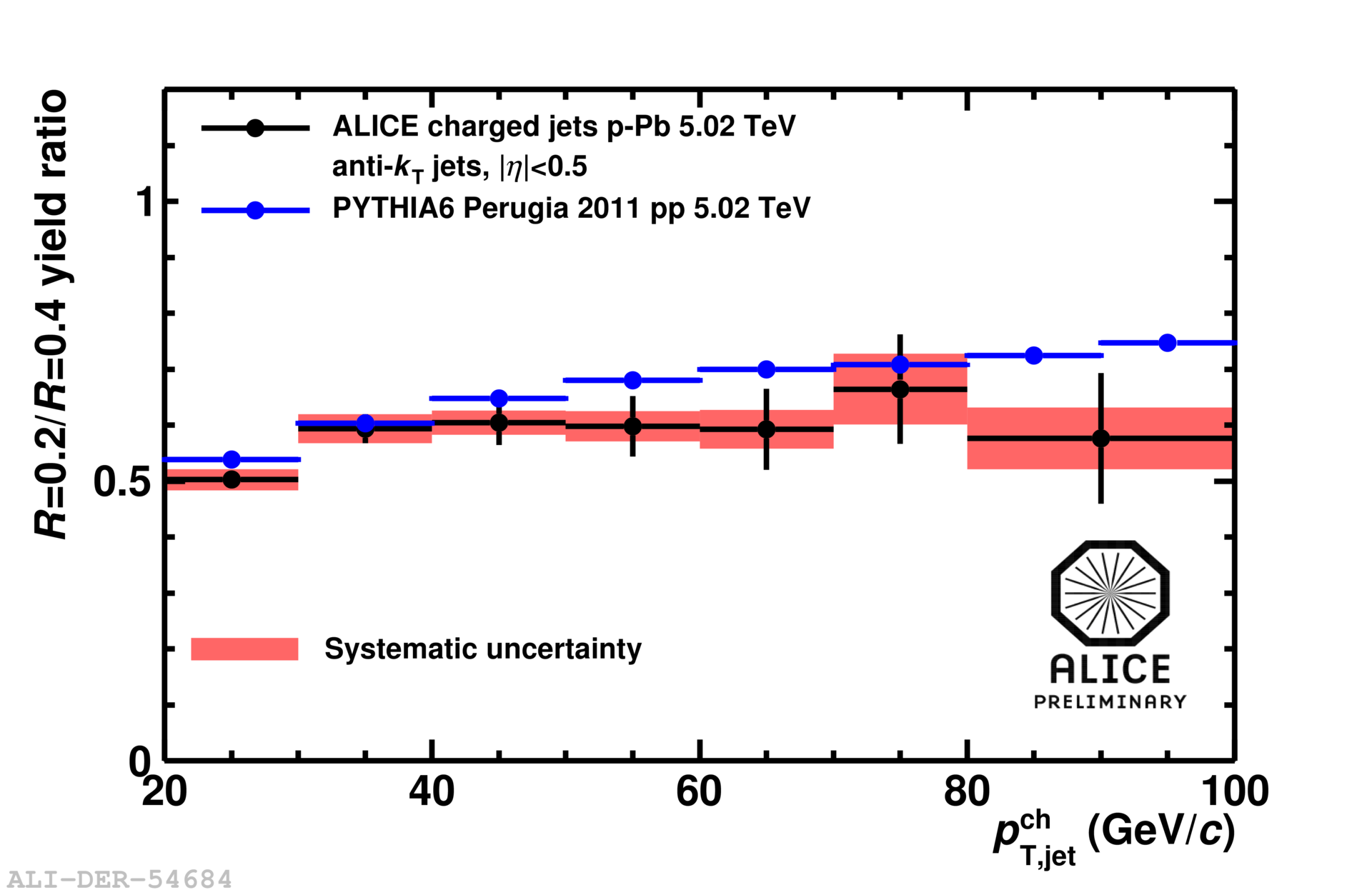}}\hfill
\caption{Charged jet yield ratio for different resolutions parameters. The data in p-Pb are compared to PYTHIA predictions at the same energy.}
\label{FinalRatioPerugia}
\end{figure}

\begin{figure}[!htp]
\centering
\subfigure{\includegraphics[width=0.81\textwidth]{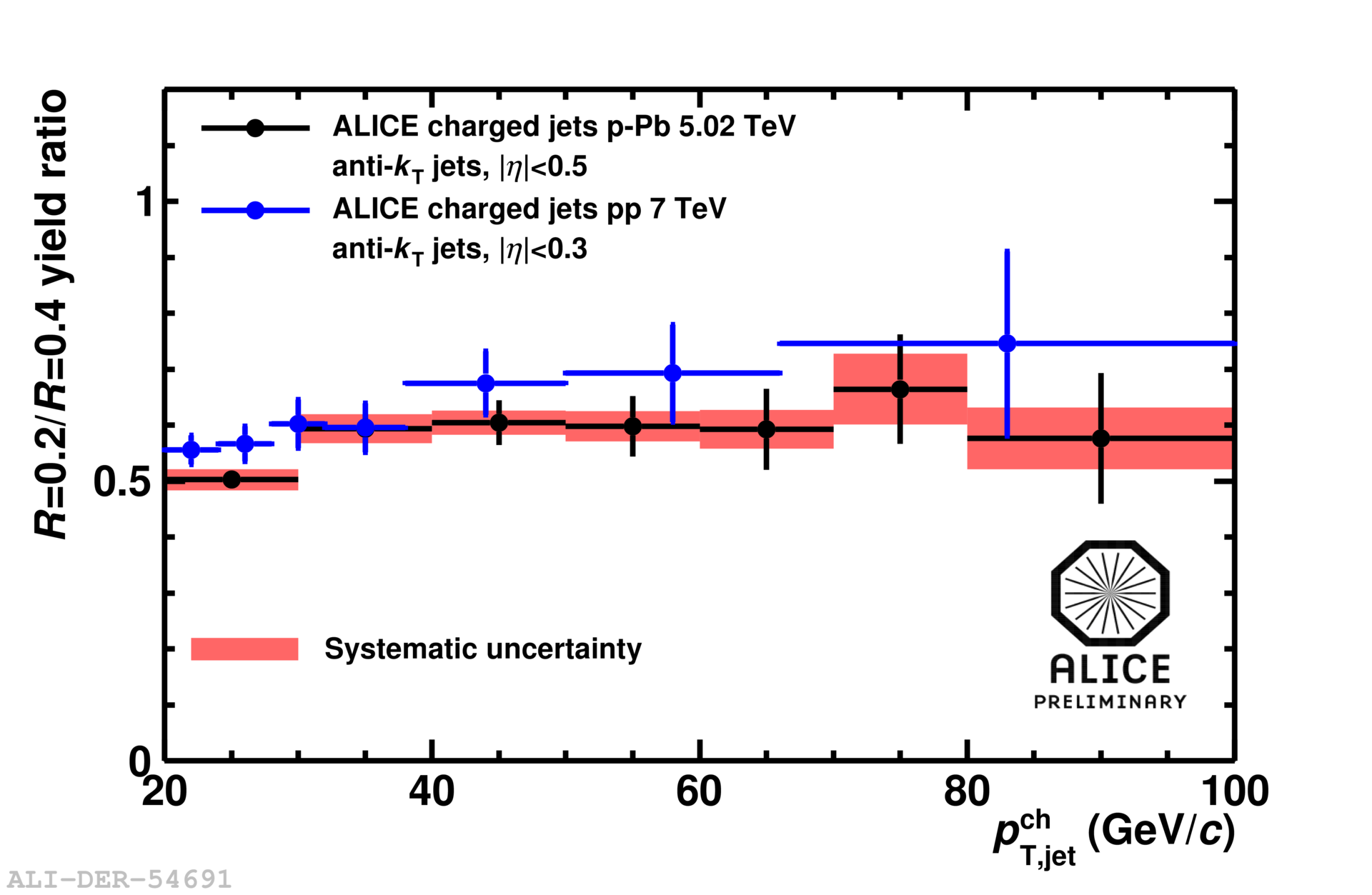}}\hfill
\caption{Charged jet yield ratio at $\sqrt{s} = 5.02$ TeV compared to the same ratio in pp at 7 TeV.}
\label{FinalRatio7TEV}
\end{figure}

Using the analysis techniques presented in the previous section, three quantities were calculated.
The charged jet spectrum in p-Pb -- the per-event transverse momentum density distribution of the jets -- as the basic observable,
\eq{eqSpectrum}{\frac{\mathrm{d}^{2}N_\mathrm{jets}}{(N_\mathrm{event} \mathrm{d}p_{T}\mathrm{d}\eta)},}
is presented in figure \ref{FinalYield} for resolution parameter $R=0.4$. Based on this measurement and the extrapolated pp reference at $5.02$ TeV, the nuclear modification factor $R_\mathrm{pPb}$ as a measure for cold nuclear matter effects,
\eq{eqRpPb}{R_\mathrm{pPb} =\frac{\tn{pPb yield}}{\tn{pp x-section}} \cdot \frac{1}{T_{\mathrm{pPb}}},}
has been extracted and is shown in figure \ref{FinalRpPb}. The $T_\mathrm{pPb}$ accounts for the increased "parton luminosity" in p-Pb. It is related to the number of binary collisions via $N_\mathrm{coll} = T_\mathrm{pPb} ~~ \sigma^\mathrm{inel. pp}$ and was calculated using the Glauber model.\\
To analyze possible cold nuclear matter effects on the jet profile, the ratio of jet yields for two different resolution parameters $R=0.2$ and $R=0.4$ in p-Pb collisions was calculated with
\eq{eqJetYieldRatio}{R = \frac{\tn{pPb yield for } R=0.2}{\tn{pPb yield for } R=0.4}}
and this ratio is compared with PYTHIA Perugia 2011 simulations in figure \ref{FinalRatioPerugia}, showing the expected collimation of the jets with increasing jet transverse momentum. In figure \ref{FinalRatio7TEV}, the yield ratio is also compared with a similar measurement in 7 TeV pp collisions.

\newpage

\newpage

\section{Conclusions}
In this paper charged jet spectra in p-Pb collisions at $\sqrt{s_\mathrm{NN}} = 5.02$ TeV with a resolution parameter $R = 0.4$ were shown up to $p_{\mathrm{T, jet}}^\mathrm{ch}$ of 120 $\mathrm{GeV}/c$. The nuclear modification factor $R_\mathrm{pPb}$ show no strong nuclear effects of the jet spectra -- it is even compatible with no effect. The jet yield ratio of $R=0.2/0.4$ is compatible with 7 TeV pp data and also with the predictions from PYTHIA Perugia 2011 at the same energy. There is no indication that there is a nuclear modification of the jet structure between $R = 0.2$ and $0.4$. Additionally, results on full jets, jet triggered data samples, and also centrality dependent quantities in p-Pb can be expected soon from ALICE.

\end{document}